\documentclass[a4paper,11pt]{article}
\usepackage[utf8]{inputenc}
\usepackage{pos}
\usepackage{cleveref}
\usepackage{siunitx}
\DeclareSIUnit \parsec {pc}
\DeclareSIUnit \GeV {GeV}
\DeclareSIUnit \TeV {TeV}
\usepackage[margin=10pt]{subcaption}
\usepackage{lineno}
\usepackage{setspace}
\usepackage{wrapfig}
\usepackage{hyperref,url}
\newcommand{\atel}[1]{\href{https://www.astronomerstelegram.org/?read=#1}{ATel #1}}
\newcommand{\gcn}[1]{\href{https://gcn.nasa.gov/circulars/#1}{GCN #1}}

\title{Low-latency neutrino follow-up combining diverse IceCube selections}

\ShortTitle{Multi-selection Fast Response}

\author{The IceCube Collaboration \\{\normalsize \normalfont(a complete list of authors can be found at the end of the proceedings)}\\}

\emailAdd{chraab@icecube.wisc.edu}
\emailAdd{sahori@icecube.wisc.edu}
\emailAdd{ssclafani@icecube.wisc.edu}
\emailAdd{jthwaites@icecube.wisc.edu}

\abstract{

Neutrino observations are a crucial component of multi-messenger astronomy, but are currently limited by effective area and high atmospheric background. However, while other telescopes with limited field of view must be pointed in order to capture observations, IceCube’s full-sky field of view and high uptime make it an excellent instrument for realtime follow-up of astrophysical transient sources.
IceCube searches for neutrino transients using an unbinned maximum likelihood method with parameters for the source’s emission time period, extension, and energy spectrum. This Fast Response Analysis can provide analysis results within tens of minutes of an astrophysical transient. Besides the follow-up of astrophysical transients manually selected as candidates, it also routinely scans areas of the sky compatible with gravitational wave alerts from LIGO/Virgo/KAGRA and IceCube event singlets which have a high probability of originating from an astrophysical source.
Currently the analysis uses TeV muon neutrino candidate events whose track signature is especially suited for a precise angular reconstruction, selected and reconstructed at the South Pole and transmitted with low-latency over a satellite connection.
Recently, IceCube and the neutrino astronomy community are evolving to use event samples constructed with different selections. These efforts include the follow-up of gravitational wave events with GeV neutrinos detected by IceCube-DeepCore and the observation of the Galactic plane with cascade events produced by all neutrino flavors.
With plans to make IceCube-DeepCore GeV neutrino candidates and cascade events available on a day-scale latency, they can also be used in Fast Response Analyses. Moreso, multiple event samples can be combined in a Fast Response Analysis that is sensitive to a broader energy range of a neutrino transient spectrum and ensures the inclusion of all neutrino flavors.
We present the analysis method and technical aspects of such an extension of the existing framework.
This includes a proposed new pipeline allowing the inclusion of the more computationally-intensive reconstruction methods used by the aforementioned event selections.
The extension is validated using example analyses implemented in this framework. 

\vspace{4mm}

{\bfseries Corresponding authors:}
Christoph Raab$^{1*}$, 
Sam Hori$^{2}$, 
Steve Sclafani$^{3}$, 
Jessie Thwaites$^{2}$\\
{$^{1}$ \itshape CP3, Université catholique de Louvain}\\
{$^{2}$ \itshape Dept. of Physics and WIPAC, University of Wisconsin-Madison}\\
{$^{3}$ \itshape University of Maryland}\\[4mm]
$^*$ Presenter}

\ConferenceLogo{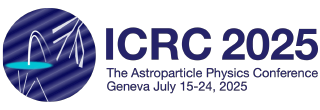}

\FullConference{39th International Cosmic Ray Conference (ICRC2025)\\
 15–24 July 2025\\
Geneva, Switzerland\\}

\begin{document}
\maketitle

\section{Motivation}
\label{sec:mot}

Many astrophysical transients, such as gamma ray bursts, supernovae, AGN flares and novae are expected to produce high-energy neutrinos via hadronic processes. The understanding of these sources would be greatly enhanced by observing such multi-messenger signals. IceCube is a cubic kilometer neutrino telescope installed at the South Pole and prominently positioned to contribute to these observations thanks to its high uptime and full-sky field of view. A small number of individual neutrino point sources associated with known counterparts has been found in archival analyses~\cite{TXS:ps,IceCube:2022der}. After an astrophysical transient candidate, providing neutrino source detections with low latency to other telescopes would permit rapid follow-up observations especially for those with a narrower field of view. In this way, a low-latency neutrino source search contributes to the characterization of the transient in question.

\section{Current usage and implementation}
\label{sec:cur}

IceCube instruments a cubic kilometer of glacial ice at the South Pole with an array of 5,160 photo-multipliers. They detect Cherenkov light induced by charged particles exceeding the speed of light in the ice. For the charged-current interactions of $\nu_\mu$ in this volume, the Cherenkov light traces the path of the outgoing muon in a so-called \emph{track} event signature. This is particularly suited for %
event reconstruction methods to infer the neutrino arrival direction. The majority of neutrino interactions meanwhile produce charged particles which lose their energy within a smaller area. This spherical signature, called a \emph{cascade}, also allows directional reconstruction, albeit with limited precision. 

While the majority of IceCube data processing happens in the northern hemisphere, during IceCube data-taking, a sub-set of events are selected and reconstructed at the South Pole~\cite{IceCube:detectorpaper, IceCube:gfu2016}. This event selection, called the Gamma-ray Follow-up (GFU) dataset, consists of high energy tracks, transmitted to the North with high priority over a low-latency, low-bandwidth satellite link~\cite{IceCube:detectorpaper} and used in several online analyses including the gamma-ray follow-up program~\cite{Mancina:2025:NewPublicNeutrino}. GFU events have a good angular resolution ($\lesssim \ang{1}$ radius) and are reconstructed with low latency.
These events are received in the northern hemisphere after roughly \qty{60}{s} and can be quickly queried from a database according to the desired follow-up time window. Together with archival data of past seasons to represent the background and Monte Carlo simulation to represent the signal, these are used in low-latency likelihood analyses. Internal reports with detector status quantities and analysis results are automatically generated as part of the same pipeline.

\begin{figure}
    \centering
    \includegraphics[width=0.6\columnwidth]{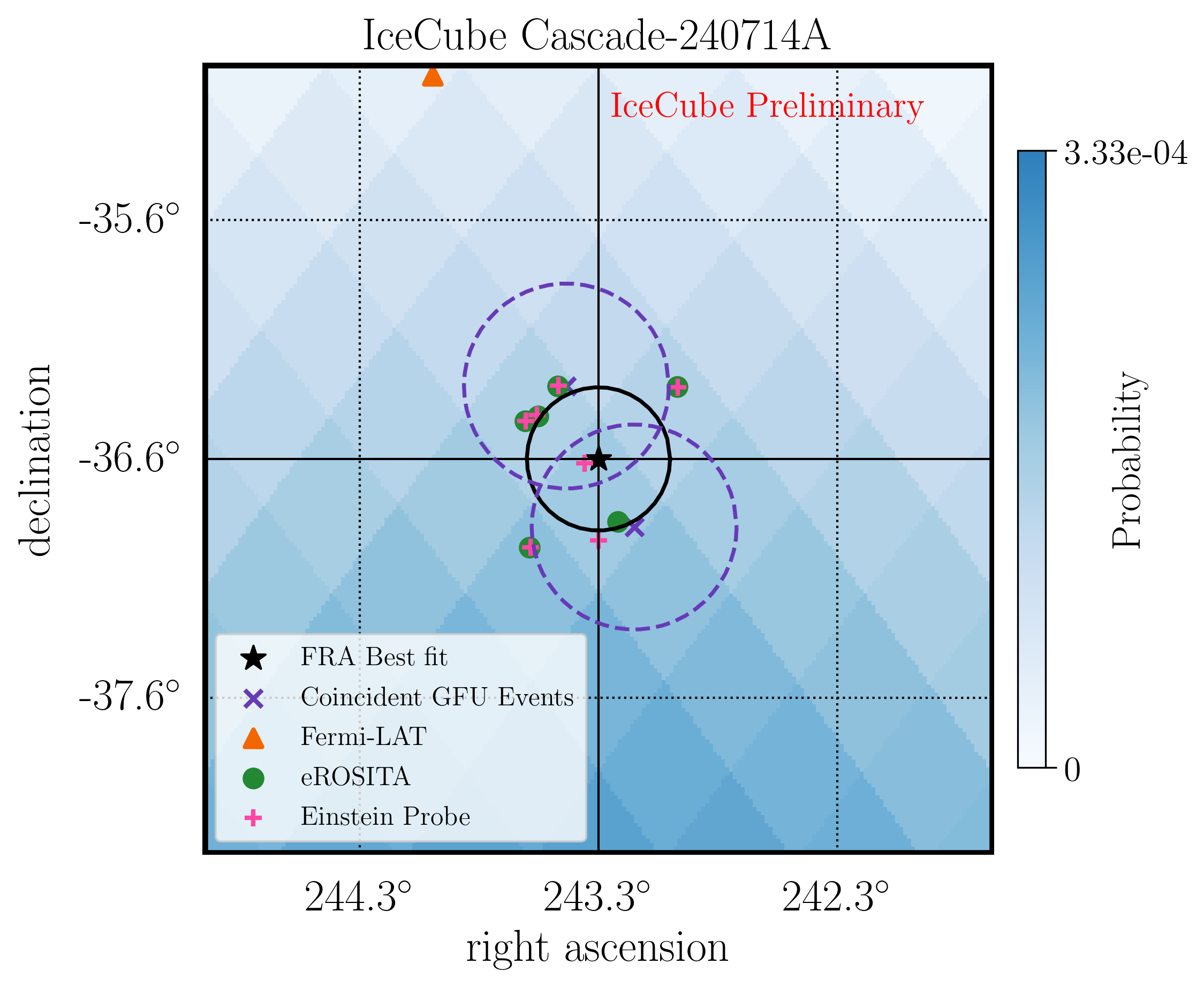}
    \caption{Skymap zoomed around the FRA best-fit for the \SI{\pm1}{day} search for IceCube-Cascade 240714A. The two coincident GFU events are shown (purple) with the FRA best-fit provided in real-time (star). In real-time, an error around the best-fit was also provided with a radius of \ang{0.3} (shown as a black circle). The nearest Fermi-LAT source (top of plot) is shown in an orange triangle. Einstein Probe followed up our ATel in real-time, and reported sources they detected (labeled Einstein Probe) and associated eROSITA catalog sources.}  
    \label{fig:cur}
\end{figure}

The current implementation of this Fast Response Analysis (FRA) is used to search for neutrino emission from a variety of transients, including gravitational wave events \cite{Marka:2025icrc} and electromagnetically-detected transients \cite{Abbasi:2021:FollowupAstrophysicalTransientsa}. The FRA is also used to search for additional track events surrounding all public IceCube alert events, which would be indicative of a neutrino flare from a source~\cite{Abbasi:2023:ConstraintsPopulationsNeutrino}. One such follow-up of an IceCube alert was the case of IceCube-Cascade 240714A (\atel{16708}). %
The FRA identified two track-like events spatially coincident with one another and the cascade alert event skymap (shown in figure \ref{fig:cur}). This represents a pre-trials p-value of $0.007$ in the analysis. The public telegram sent by IceCube saw follow-ups from multiple observatories, including Fermi-LAT (\gcn{36892}) and Einstein Probe (EP, \gcn{36894}). Fermi-LAT saw no excess gamma-ray emission consistent with the best-fit position. EP identified 7 sources, including two within the \ang{0.3} error circle of the analysis best-fit, one of which was previously uncatalogued. The EP sources were followed up by the Thai Robotic Telescope network, which found no uncatalogued sources consistent with the EP sources, and no apparent brightening for cataloged sources in the R-band (\gcn{36902}). This example illustrates the interest of the community in these short timescale analyses performed in the FRA, in order to rapidly detect and observe potential multimessenger sources.

\section{Reduced-latency data stream method}
\label{sec:dat}

The GFU track events provide good spatial resolution, but rejecting the background of atmospheric muons induced by cosmic rays requires a higher energy threshold in the southern sky. Transient neutrino source searches are expanding to use more diverse event selections. This includes for example cascade events~\cite{BalagopalV:2025:MultiEnergyMultiSampleSearches}, whose event signature is distinct from atmospheric muons, allowing for a more even sky coverage of all neutrino transient candidates. The DNN Cascades selection in particular was the sample used in the observation of a $4.5\sigma$ excess from the Galactic plane~\cite{IceCube:2023ame}.
IceCube further possesses a denser in-fill array called DeepCore, allowing it to extend to sub-TeV energies, which is well-motivated  by the predicted emission of GeV neutrinos from e.g. GRBs~\cite{Murase:2022:NeutrinosBrightestGammaRaya}. Available event selections are e.g. the reconstructed \textsc{GRECO}\footnote{GeV Reconstructed
Events with Containment for Oscillation} tracks and cascades~\cite{BalagopalV:2025:MultiEnergyMultiSampleSearches}. The \textsc{ELOWEN}\footnote{Extremely LOW ENergy} selection contains the faintest events triggering DeepCore ~\cite{Lamoureux:2025:ProbingNeutrinoEmission}, but currently can only rely on timing information to identify astrophysical activity. The goal of the current work is to allow a low-latency application of these event selections to transient searches.

\begin{figure}
    \centering
    \includegraphics[width=0.98\linewidth]{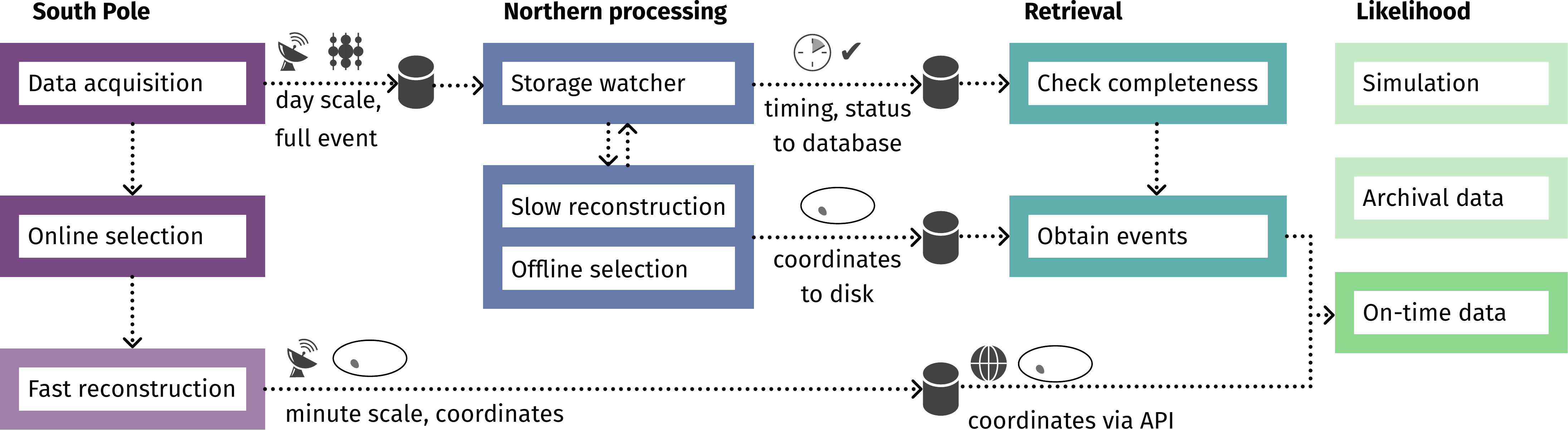}
    \caption{Flow chart of reduced latency data stream methods. The upper path corresponds to the day-scale latency method proposed in \cref{sec:dat} for DNN Cascades and GRECO, while the lower path describes the one currently used by GFU.}
    \label{fig:dat/flow}
\end{figure}

Some of these new event selections use more computationally intensive reconstruction methods, which unlike GFU's can not be run at the South Pole in tandem with data acquisition. We propose an approach, illustrated by the diagram in \cref{fig:dat/flow}, to nevertheless make these accessible with day-scale latency, allowing for a range of follow-up analyses analogous to the archival ones.

Event data is transmitted from the South Pole over a high-bandwidth partial-uptime satellite link~\cite{IceCube:detectorpaper} and arrives in the North typically within $\approx16$ hours (half of events) to $\approx1.5$ days ($>99$\%), divided into $\mathcal{O}(1000)$ files per calendar day.
Each file is bundled with metadata indicated the period of data covered therein. As it is forwarded to event selection and reconstruction, this metadata is stored in a tracking database. Upon completion, the processing success is stored in the same database and the reconstructed events in a separate repository, indexed by calendar day for convenient retrieval of on-time events. 

\begin{wrapfigure}{r}{0.5\linewidth}
    \centering
    \includegraphics[width=\linewidth]{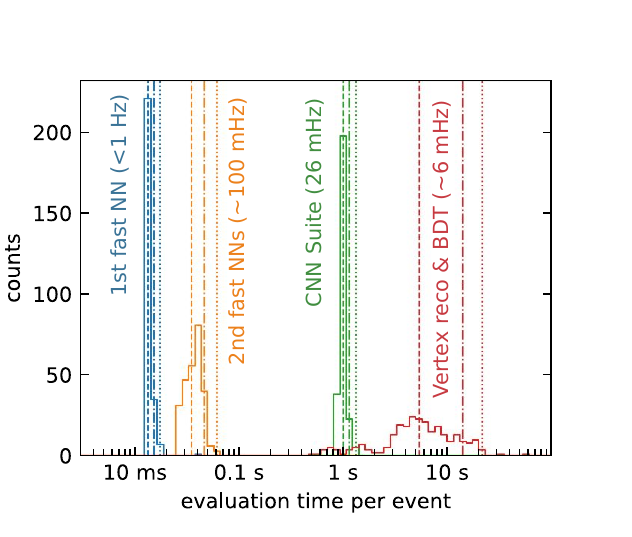}
    \caption{CPU time required per event for stages of the DNN Cascades selection leading up to the reconstruction. As the rate decreases, more computationally intensive methods can be used.}
    \label{fig:dat/cpu}
\end{wrapfigure}

The DNN Cascade selection uses a series of neural networks (NNs) and boosted decision trees (BDTs) to quickly identify cascade-like events in IceCube and reject incoming atmospheric muons. The selection is trained on data and simulation using summarized information based on the event light deposition. This differs from the previous strategy of using high-level information.  Due to the speed of the NNs and the removal of high-level reconstruction, the selection can be applied at an earlier stage of the data selection pipeline. The result is a dataset that improves the efficiency when compared to previous selections. Between 2011 and 2021, 6\% of its events are atmospheric muons, and 87\% atmospheric neutrinos.  The remaining 7\% of events are attributed to astrophysical neutrinos~\cite{IceCube:2023ame}.  

These events have an angular resolution that is $\sim10^\circ$ at energies above \qty{10}{TeV}, compared to GFU reaching sub-degree resolution.  Despite the larger uncertainty, the purity of the dataset and the improved sensitivity in the southern hemisphere make realtime or near-realtime followup with this selection particularly interesting.  
The final reconstruction takes 2--3 minutes to complete for most events.%

IceCube can send compressed event data records to the North with low latency at a limited rate. 
This can eventually be used in a scheme that runs the first stage of the DNN Cascades selection at the South Pole, reducing the data rate to \SI{6}{mHz}, shown in \cref{fig:dat/cpu}. Event data at this rate can then be transmitted to the North, and after applying the final reconstruction and selection with minimum latency these events would also available for analysis within tens of minutes.

Both the day-scale and minute-scale processing methods can be applied to GRECO, introduced previously, which currently also is reconstructed in the North with dedicated methods to accommodate for the low light yield in these events. These take an average of 2 minutes to complete, although some may take up to an hour, emphasizing the importance of asynchronous processing.

\section{Data quality monitoring}
\label{sec:mon}

\begin{wrapfigure}{r}{0.4\columnwidth}
    \centering
    \includegraphics[width=\linewidth]{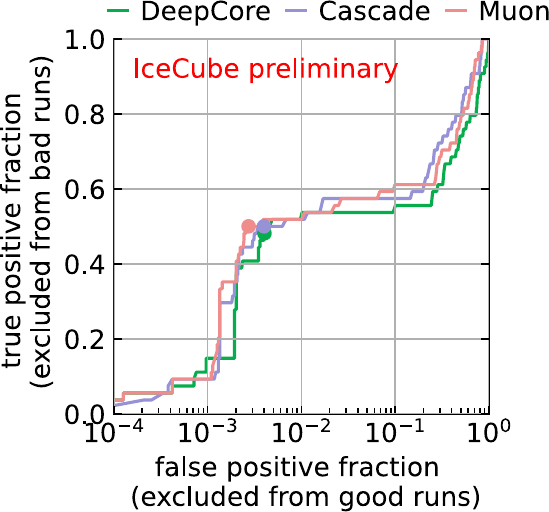}
    \caption{ROC curves comparing possible thresholds to be set on the three instability scores. The x-axis is a "false positive" rejection rate during good runs, while the y-axis is the "true positive" rejection during bad runs. The threshold of 10 is indicated by filled circles on the respective curves.}
    \label{fig:mon/roc}
\end{wrapfigure}

IceCube data-taking proceeds in periods called \emph{runs} which have a typical duration of 8~hours. The condition of the detector during one run and resulting data quality are verified manually afterwards, with help of an array of monitored quantities and knowledge of past problems. This allows to define the set of runs usable for a self-consistent archival data set. For FRA however, a heuristic is required that can assure data quality with a reasonable degree of confidence with lower latency. This has already been implemented to accommodate the GFU event selection~\cite{IceCube:gfu2016}. This heuristic's starting point are the rates of intermediate event selection stages, measured in time intervals of typically 600 seconds. With this rate as $X$ and the exponentially weighted averages $\langle X \rangle$ and $\langle X^2 \rangle$, a Z-score is calculated for the deviation of $X - \langle X \rangle$ relative to the standard deviation.

Summing this Z-score for several rates as well as the ratio between them results in the final instability score, which can be compared to a threshold. We present here instability scores that are analogously derived for a cascade selection like DNN Cascades, and a DeepCore-focused selection like \textsc{GRECO} or \textsc{ELOWEN}. These might eventually prove to take complementary roles to the original, ensuring the quality of the specific event selections.

In the new definitions, 
the Z-scores are also weighted proportionally to $\sqrt{\Delta t}$ in accordance with the expected variance. This suppresses the statistical fluctuations present for the GFU instability score during shorter-than-typical bins.
\Cref{fig:mon/roc} shows that both new scores remain sensitive to conditions flagged by the run monitoring, and a common threshold or multi-variate cut can be optimized for a particular analysis.

\section{Examples and performance}
\label{sec:ana}

\begin{figure}
    \centering
    \begin{subfigure}[t]{0.48\linewidth}
    \includegraphics[width=1\columnwidth]{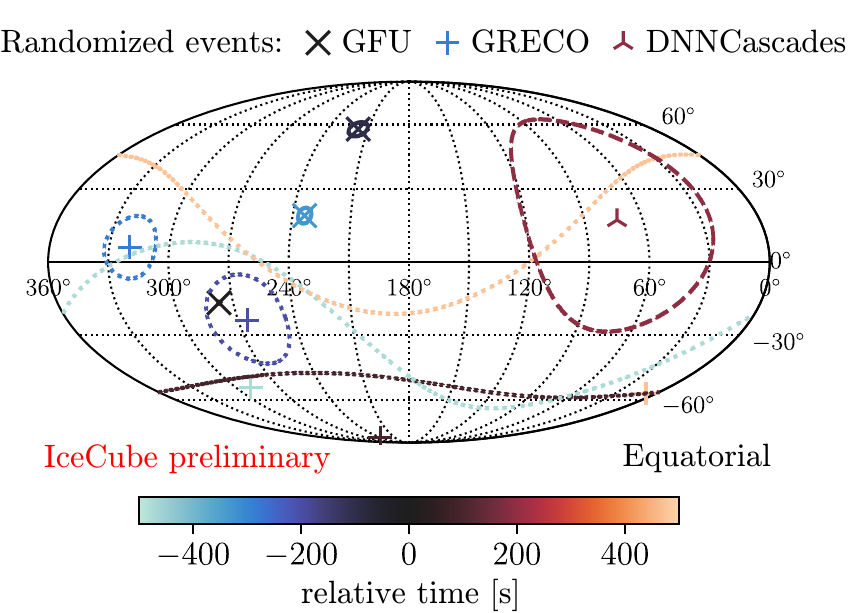}
    \caption{Sky map of randomized events within a 1000-second interval including one DNN Cascades event.}
    \label{fig:ana/scatter}
    \end{subfigure}
    \begin{subfigure}[t]{0.48\linewidth}
    \includegraphics[width=1\columnwidth]{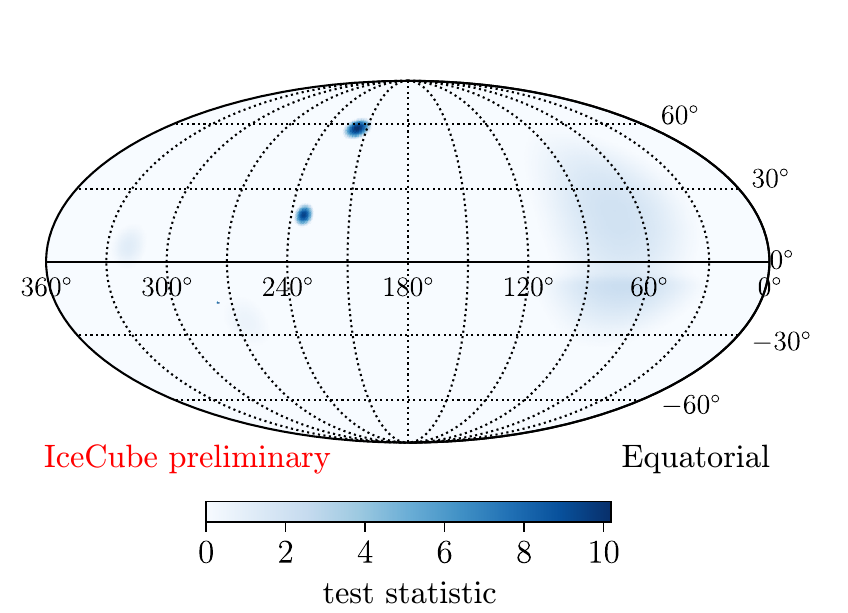}
    \caption{Corresponding test statistic map in a point source analysis combining the three event selections.}
    \label{fig:ana/ts}
    \end{subfigure}
    \caption{}
    \label{fig:ana/sky}
\end{figure}

As an example we show the combination of three event selections in a point source search, using the  previously implemented unbinned likelihood method~\cite{IceCube:ps7year}. 
Instead of fitting the source hypothesis with each selection independently, they are able to share its parameters: position, time window, spectral index and flux normalization. Each event selection's acceptance to this signal determines the expected number of detected events. %

The selections comprise the GFU high-energy tracks (already used in the current implementation), the reconstructed DeepCore tracks and cascades of \textsc{GRECO}, and the high-energy cascades in DNN Cascades. The use of these selections for neutrino follow-up mirrors the original FRA~\cite{Abbasi:2021:FollowupAstrophysicalTransientsa} as well as archival analyses of gravitational wave events~\cite{BalagopalV:2025:MultiEnergyMultiSampleSearches}.
We include %
events that appear in multiple event selections: 1.8\% of GFU events are also in \textsc{GRECO} and 0.03\% also in DNN Cascades, and the latter selections have an overlap of 0.2\%.

A sky map of a simulated background-only observation in a \SI{1000}{s} window follow-up is shown in \cref{fig:ana/scatter}. Scanning the full sky, these then correspond to the test statistic map in \cref{fig:ana/ts}, which the FRA can combine with the probability sky map of a neutrino or gravitational wave event~\cite{Abbasi:2021:FollowupAstrophysicalTransientsa} for a joint search and possible improved localization.%

\begin{figure}
    \centering
    \begin{subfigure}[t]{1\linewidth}
    \includegraphics[width=0.9\columnwidth]{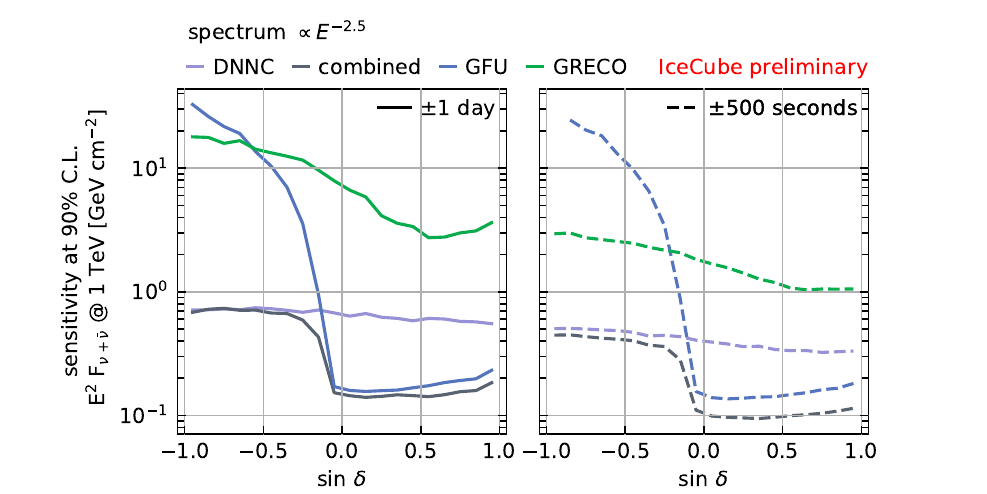}
    \caption{Time-integrated sensitivity towards a power-law $\nu+\bar{\nu}$ flux, per-flavor at \qty{1}{\TeV}. Shown are follow-ups in time windows of $\pm\qty{1}{d}$ (solid lines) and $\pm\qty{500}{s}$ (dashed lines). Each is performed individually with one of three event selections: \textsc{GRECO} (green), GFU (blue), DNN Cascades (purple) or their multi-FRA combination (gray).}
    \label{fig:ana/dec}
    \end{subfigure}
    \begin{subfigure}[t]{\linewidth}
    \includegraphics[width=0.9\columnwidth]{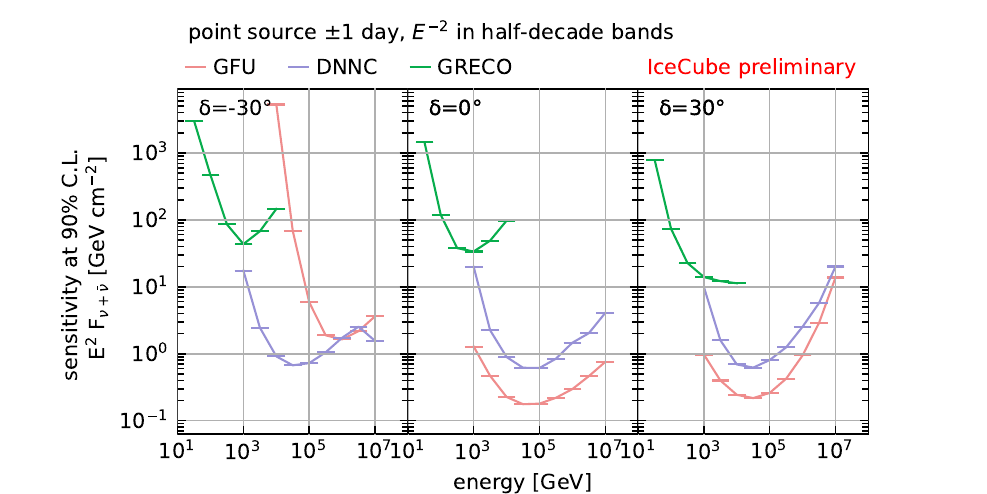}
    \caption{Differential sensitivity in energy for an $E^{-2}$ per-flavor $\nu+\bar{\nu}$ flux in half-decade energy bands, assuming a time window of \qty{\pm1}{day}. Each is performed individually with one of three event selections: GRECO (green), GFU (red), DNN Cascades (purple).}
    \label{fig:ana/diff}
    \end{subfigure}
    \caption{}
    \label{fig:ana/sens}
\end{figure}

Sensitivities are calculated at 90\% C.L. for a search in commonly used time windows of $\qty{\pm500}{s}$ or $\qty{\pm1}{day}$. 
The comparison in \cref{fig:ana/dec} shows that addition of other event selections to GFU helps regain sensitivity in the southern hemisphere lost to the latter's muon background.
The number of background events within a typical point spread function meanwhile affects the sensitivity more for follow-ups in longer time windows. 
We also show differential sensitivities versus neutrino energy, which show the complementarity of event selections in \cref{fig:ana/diff}.

\section{Conclusions and outlook}
\label{sec:out}

Building on a powerful framework and program (see \cref{sec:cur}), the presented work shows a first step towards more comprehensive neutrino follow-ups which cover a larger fraction of the neutrino transient hypothesis space (see \cref{sec:ana}). This involves a more even coverage of the sky, as well as sensitivity to GeV neutrino emission, shown in \cref{fig:ana/sens}. The implementation supports arbitrary event selections. %
In the future, the latency will be further decreased by implementation of the minute-scale approach described in \cref{sec:dat}. Remaining points to resolve include the removal of events appearing in multiple event selections, although this is a rare exception.

Already now, the multi-selection FRA can accommodate spectral parameters, although so far it has only been used with single power-law spectra.  Incorporating multiple independent spectral components promises to make fuller use of the broadened energy range. At extremely low energies, events are no longer reconstructed the same way and require a different formulation to the corresponding likelihood term, which remains for future work. %

\bibliographystyle{bib/JHEP}
\begingroup
\bibliography{bib/references,bib/zotero,bib/icrc2025}
\endgroup

\clearpage

\section*{Full Author List: IceCube Collaboration}

\scriptsize
\noindent
R. Abbasi$^{16}$,
M. Ackermann$^{63}$,
J. Adams$^{17}$,
S. K. Agarwalla$^{39,\: {\rm a}}$,
J. A. Aguilar$^{10}$,
M. Ahlers$^{21}$,
J.M. Alameddine$^{22}$,
S. Ali$^{35}$,
N. M. Amin$^{43}$,
K. Andeen$^{41}$
% ...
\\
\\
% ...
$^{10}$ Universit{\'e} Libre de Bruxelles, Science Faculty CP230, B-1050 Brussels, Belgium \\
% ...
$^{16}$ Department of Physics, Loyola University Chicago, Chicago, IL 60660, USA \\
$^{17}$ Dept. of Physics and Astronomy, University of Canterbury, Private Bag 4800, Christchurch, New Zealand \\
% ...
$^{21}$ Niels Bohr Institute, University of Copenhagen, DK-2100 Copenhagen, Denmark \\
$^{22}$ Dept. of Physics, TU Dortmund University, D-44221 Dortmund, Germany \\
% ...
$^{35}$ Dept. of Physics and Astronomy, University of Kansas, Lawrence, KS 66045, USA \\
% ...
$^{39}$ Dept. of Physics and Wisconsin IceCube Particle Astrophysics Center, University of Wisconsin{\textemdash}Madison, Madison, WI 53706, USA \\
% ...
$^{41}$ Department of Physics, Marquette University, Milwaukee, WI 53201, USA \\
% ...
$^{43}$ Bartol Research Institute and Dept. of Physics and Astronomy, University of Delaware, Newark, DE 19716, USA \\
% ...
$^{63}$ Deutsches Elektronen-Synchrotron DESY, Platanenallee 6, D-15738 Zeuthen, Germany \\
$^{\rm a}$ also at Institute of Physics, Sachivalaya Marg, Sainik School Post, Bhubaneswar 751005, India \\
% ...

\subsection*{Acknowledgments}

\noindent
The authors gratefully acknowledge the support from the following agencies and institutions:
USA {\textendash} U.S. National Science Foundation-Office of Polar Programs,
U.S. National Science Foundation-Physics Division,
U.S. National Science Foundation-EPSCoR,
U.S. National Science Foundation-Office of Advanced Cyberinfrastructure,
Wisconsin Alumni Research Foundation,
Center for High Throughput Computing (CHTC) at the University of Wisconsin{\textendash}Madison,
Open Science Grid (OSG),
Partnership to Advance Throughput Computing (PATh),
Advanced Cyberinfrastructure Coordination Ecosystem: Services {\&} Support (ACCESS),
Frontera and Ranch computing project at the Texas Advanced Computing Center,
U.S. Department of Energy-National Energy Research Scientific Computing Center,
Particle astrophysics research computing center at the University of Maryland,
Institute for Cyber-Enabled Research at Michigan State University,
Astroparticle physics computational facility at Marquette University,
NVIDIA Corporation,
and Google Cloud Platform;
Belgium {\textendash} Funds for Scientific Research (FRS-FNRS and FWO),
FWO Odysseus and Big Science programmes,
and Belgian Federal Science Policy Office (Belspo);
Germany {\textendash} Bundesministerium f{\"u}r Forschung, Technologie und Raumfahrt (BMFTR),
Deutsche Forschungsgemeinschaft (DFG),
Helmholtz Alliance for Astroparticle Physics (HAP),
Initiative and Networking Fund of the Helmholtz Association,
Deutsches Elektronen Synchrotron (DESY),
and High Performance Computing cluster of the RWTH Aachen;
Sweden {\textendash} Swedish Research Council,
Swedish Polar Research Secretariat,
Swedish National Infrastructure for Computing (SNIC),
and Knut and Alice Wallenberg Foundation;
European Union {\textendash} EGI Advanced Computing for research;
Australia {\textendash} Australian Research Council;
Canada {\textendash} Natural Sciences and Engineering Research Council of Canada,
Calcul Qu{\'e}bec, Compute Ontario, Canada Foundation for Innovation, WestGrid, and Digital Research Alliance of Canada;
Denmark {\textendash} Villum Fonden, Carlsberg Foundation, and European Commission;
New Zealand {\textendash} Marsden Fund;
Japan {\textendash} Japan Society for Promotion of Science (JSPS)
and Institute for Global Prominent Research (IGPR) of Chiba University;
Korea {\textendash} National Research Foundation of Korea (NRF);
Switzerland {\textendash} Swiss National Science Foundation (SNSF).

\end{document}